\documentclass[a4paper,UKenglish,cleveref,autoref,thm-restate]{lipics-v2021}

\hideLIPIcs 

\usepackage[noend, boxed]{algorithm2e}
\usepackage{booktabs}
\usepackage[sort,nocompress]{cite} 

\newif\ifreviewnotes
\reviewnotesfalse
\ifreviewnotes
  \usepackage{todonotes}
\else
  \usepackage[disable]{todonotes}
\fi

\newcommand{\Oish}{\widetilde{O}}
\newcommand{\Omegaish}{\widetilde{\Omega}}

\newcommand{\dist}{\texttt{dist}}

\newcommand{\PathBFS}{\texttt{PathBFS}}

\title{On the Communication Complexity of Shortcut Set Computation} 

\titlerunning{Communication Complexity of Shortcut Set Computation}

\author{Gregory Bodwin}{University of Michigan, USA}{}{}{}

\author{Luba Samborska}{University of Michigan, USA}{}{}{}

\authorrunning{G. Bodwin and L. Samborska} 

\Copyright{Gregory Bodwin and Luba Samborska} 



\nolinenumbers 

\EventLongTitle{44th International Symposium on Theoretical Aspects of Computer Science (STACS 2027)}
\EventShortTitle{STACS 2027}
\EventAcronym{STACS}
\EventYear{2027}
\EventDate{March 8--12, 2027}
\EventLocation{G\"ottingen, Germany}
\SeriesVolume{}
\ArticleNo{}

\begin{document}

\maketitle

\begin{abstract}
A recent breakthrough of Kogan and Parter (SODA '22) provided the first polynomial improvement in decades on the size/diameter tradeoff available for shortcut sets.
However, algorithmic applications of shortcut sets have mostly not taken advantage of these new bounds.
A barrier is that many shortcut set applications occur in \emph{decentralized} settings, and we generally lack decentralized algorithms implementing the Kogan-Parter bounds.

To this end, we study the \emph{communication complexity} of constructing Kogan-Parter shortcut sets.
The edges of an $n$-node directed graph are split between two players, Alice and Bob, who are also given a target diameter $D$.
Our main result is a protocol that, for $D \le n^{1/3}$, allows Alice and Bob to agree on a shortcut set matching the Kogan-Parter tradeoff between size and diameter, while exchanging only $\Oish(n^2 / D^2)$ bits of communication.
We also provide (non-matching) lower bounds on the communication complexity required to find a correct $D$-diameter shortcut set of any size.

The main technical ingredient for the upper bounds is a low-communication protocol for finding a \emph{path cover}, a common subroutine in shortcut set constructions that has attracted some independent interest.
Our protocol crucially avoids applications of Menger's theorem or antichain decompositions, as often done in prior work.
On the lower bounds side, we relate the communication complexity of the shortcut set problem to the size of bridge-free path systems. 
\end{abstract}

\clearpage 
\section{Introduction}

We study \emph{shortcut sets}, a recently-popular primitive in network design:

\begin{definition} [Shortcut Sets \cite{Thorup92}]
    Given a directed $n$-node graph $G$ with transitive closure $G^*$, a $D$-diameter reducing shortcut set is a set of edges $H \subseteq E(G^*)$ such that, for all pairs $(u,v) \in E(G^*)$, we have $\dist_{G \cup H}(u,v) \leq D.$
\end{definition}

Shortcut sets are studied in part for their algorithmic applications in various decentralized models of computation, like the parallel, distributed, or dynamic models.
In these settings, there are some popular algorithms for reachability or distance problems where the main complexity measure scales with the diameter of the input graph.
A shortcut set can be added to the graph as a preprocessing step, thus reducing complexity while leaving reachability information unchanged.
This strategy has been extensively used in recent work on reachability and shortest path computation, e.g., \cite{UY91, cohen2000, elkin2019RNC, cao_efficient_2020, cao2020improved, cao2023exact, fineman2018nearly, JLS19, gutenberg2020, henzinger2014, LN2022, elkin2019journal, elkin2017, censor2021, bernstein2021deterministic, chen2022maximum}.

\begin{table}[htbp]
\begin{center}
\begin{tabular}{ll}
\toprule
\textbf{Shortcut Set size} & \textbf{Citation}\\
\midrule
$\Oish\left(n^2 D^{-2}\right)$ & \cite{UY91} (see also \cite{Fineman19, JLS19, BGJRW12})\\
$O\left(n^2 D^{-2}\right)$ & \cite{BRR10}\\
$\Oish\left( n^{3/2} D^{-3/2} + n^2 D^{-3}  \right)$ & \cite{KP22a} (see also \cite{BBBDFN25})\\
\midrule
$\Omega\left(n^{\frac{2d}{d+1}} D^{-d}\right)$ & \cite{HP18} (see also \cite{Hesse03, BHT23})\\
$\Omegaish\left( n^{4/3} D^{-4/3} + n^{5/4} D^{-1} \right)$ & \cite{BH23}\\
$\Omega\left(n^{1 + \frac{d-1}{d+3}} D^{-(d-1)} \right)$ & \cite{VXX24}\\
\bottomrule
\end{tabular}
\end{center}
\caption{\label{tbl:priorwork} Known upper and lower bounds on the size $|H|$ of $D$-diameter reducing shortcut sets.}
\end{table}

For these applications, it is typically necessary to limit the number of edges $|H|$ in the shortcut set to (roughly) the number of edges $|E(G)|$ in the input graph, so that adding the shortcut edges does not asymptotically change $|E(G)|$.
This has motivated a long line of work studying the extremal tradeoff between the size $|H|$ and diameter bound $D$; we summarize the progression of results in Table \ref{tbl:priorwork}.
For example, for many years, essentially the best known construction was a simple folklore random sampling algorithm attributed to Ullman and Yannakakis \cite{UY91}, which guaranteed shortcut sets of size $|H| \le \Oish(n)$ that reduced diameter to $D \le \Oish(n^{1/2})$.
Although this algorithm was centralized, this size/diameter tradeoff was later recovered by various \emph{decentralized} algorithms \cite{Fineman19, JLS19, BGJRW12}, which unlocked algorithmic applications.
The first polynomial improvement on the Ullman-Yannakakis tradeoff was only achieved recently, in a 2022 breakthrough by Kogan and Parter \cite{KP22a}:

\begin{theorem} [\cite{KP22a}] \label{thm:kpintro}
Every directed $n$-node graph $G$ has a $\Oish(n^{1/3})$-diameter reducing shortcut set $H$ of size $|H| \le \Oish\left( n \right)$.
\end{theorem}


One would naturally hope to apply these new shortcutting bounds to improve algorithmic applications.
However, there is a barrier: so far the algorithms implementing Theorem \ref{thm:kpintro} have been centralized.
It is not yet clear how to efficiently construct these shortcut sets in \emph{decentralized} models, with sufficiently favorable complexity to avoid wiping out the savings from using shortcut sets in the first place.
The goal of this paper is to take a step in this direction, by implementing Theorem \ref{thm:kpintro} (and other points on the size/diameter tradeoff curve proved by Kogan and Parter \cite{KP22a}) with low \emph{communication complexity}.
In addition to intrinsic interest in communication complexity as a decentralized model, it is studied due to its extensive connections to other decentralized domains.
For an overview we refer to the lecture notes of Roughgarden \cite{Roughgarden16}; some specific applications of communication complexity in the \textsc{congest} distributed model can be found in \cite{DHKKNPPW11, BCDELP19, CKPY18, DKO14, Elkin04, FHW12, HW12, Nanongkai14, NDP11, PR00}, for example.

\subsection{Our Results}

We study the two-party communication model.
Two players, Alice and Bob, each receive a set of edges on $n$ nodes $V$. Let us call Alice's set $E_A$ and Bob's set $E_B$. As part of the input, Alice and Bob are also given a parameter $D$.
Their goal is to exchange a small number of bits of information until the players agree on a valid shortcut set $H$ for $G = (V, E_A \cup E_B)$, satisfying a given size bound for $|H|$ and reducing the diameter of the graph to $\leq D$.
Our goal will also be to guarantee that $H$ reaches the existential size/diameter guarantee proved by Kogan and Parter \cite{KP22a}.
To our knowledge, this is the first time that shortcut sets have been considered in this model of computation.
We show the following bounds:

\begin{theorem}[Main Result, Upper Bound] \label{thm:main-ub}
    For any $n$-node directed graph $G = (V, E = E_A \cup E_B)$ and any integer $1 \leq D \leq n^{1/3}$, in the two-party communication model where Alice has $E_A$ and Bob has $E_B$, there is a randomized communication protocol that (deterministically) produces an edge set $H$ of size
    $$|H| \le \Oish\left(\frac{n^2}{D^3}\right)$$
    and (deterministically) uses $\Oish(n^2 / D^2)$ bits of communication, and where $H$ is a $D$-diameter reducing shortcut set with high probability.
\end{theorem}

These shortcut sets match the size/hopbound tradeoff proved by Kogan and Parter \cite{KP22a} in the parameter range $D \le n^{1/3}$.\footnote{Kogan and Parter also gave bounds for the regime $D > n^{1/3}$, where the shortcut set will have size $|H| \ll n$.  We do not reproduce these.}
Our second main result is on the lower bound side. 




\begin{theorem}[Main Result, Lower Bound] \label{thm:main-lb}
    For any integer $c\ge 1$ and constant $0\le \rho<1/2$, there is a constant $\delta=\delta(c,\rho)>0$ such that the following holds.
    For all sufficiently large $n$ and every integer $1\le D\le \delta n$, any randomized two-party protocol that, on every input, makes both players agree on a valid $D$-diameter-reducing shortcut set with probability at least $1-\rho$ requires
$$\Omega\!\left(\frac{n^{2c/(c+1)}}{D^{c-1}}\right)$$
bits of communication in the worst case. This lower bound holds even when there is no restriction on the size of the shortcut set.
\end{theorem}

To illustrate: for the choice $D = n^{1/3}$ as in Theorem \ref{thm:kpintro}, the lower bound on bits of communication is
$$\Omegaish\left( n^{\frac{2c}{c+1} - \frac{c-1}{3}} \right).$$
This is optimized equally well by the choices $c=1$ or $c=2$, both of which lead to a lower bound of $\Omegaish(n)$ on communication.
For $D = n^{1/4}$, the lower bound is
$$\Omegaish\left( n^{\frac{2c}{c+1} - \frac{c-1}{4}} \right),$$
which is optimized by the choice $c=2$, giving an improved lower bound of $\Omegaish(n^{13/12})$.
In general, smaller functions $D$ will be optimized by larger integers $c$, and yield higher lower bounds.
The lower bound in Theorem \ref{thm:main-lb} arises from a reduction to the extremal function of dense path systems without bridge subsystems, as we overview in Section \ref{sec:lboverview}.

\subsection{Other Related Work}

Alternate greedy algorithms for shortcut sets have been investigated in \cite{berman_finding_nodate} (in the context of transitive-closure spanners, a closely related object), and in \cite{BBBDFN25}.

The earliest lower bound for $O(n)$ size shortcut sets was $\Omega(n^{1/17})$, achieved by Hesse \cite{Hesse03}. This was later improved to $\Omega(n^{1/6})$ by Huang and Pettie \cite{HP18}. The current state of the art is $\Omegaish(n^{1/4})$, achieved by Bodwin and Hoppenworth \cite{BH23}.
Recent work on \emph{certified shortcut sets} \cite{haeupler2026certified} has provided analogous lower bounds for a narrower object.

For $O(m)$ shortcut sets, the earliest established lower bound was $\Omega(n^{1/17})$ by Hesse \cite{Hesse03}. This was later improved to $\Omega(n^{1/8})$ \cite{huang_lower_2021, lu_better_2022}, and then to $\Omega(n^{1/5})$ by Vassilevska Williams, Xu, and Xu \cite{VXX24}. The current state of the art is $\Omega(n^{2/9})$ \cite{HXX25}.

Shortcut sets have also played a central role in parallel algorithms for directed reachability. Fineman used shortcut constructions to obtain a nearly work-efficient parallel algorithm for single-source reachability \cite{fineman2018nearly}, which has depth $\Oish(n^{2/3})$. Later work by Jambulapati and others improved the depth to $n^{1/2 + o(1)}$ \cite{JLS19}.

In the direction of communication complexity of network design problems, Fernandez, Woodruff, and Yasuda studied graph spanners in a message-passing model \cite{FWY20}. 


\section{Technical Overview}

\subsection{Upper Bound Approach \label{sec:ubapproach}}

The communication complexity of computing shortcut sets has not been explicitly studied previously.
However, for any diameter bound $D \leq n^{1/3}$, there is a straightforward implementation of ideas in prior work that gives an upper bound of $\Oish(n^2/D)$. 
We first sketch this straightforward upper bound. 
Then we will overview the new ideas that we introduce in order to improve the upper bound to $\Oish(n^2/D^2)$.
We will simplify our discussion by focusing on the setting $D = n^{1/3}$, but these ideas extend immediately to general $D$, which is considered in the paper. 


\subsubsection{The Simple $\Oish(n^{5/3})$-Communication Upper Bound}




First, Alice and Bob agree on a random subset $V' \subseteq V$ of $\Oish(n^{2/3})$ nodes.
Then, Alice and Bob run an out- and in-BFS search from each node in $V'$.
There is a protocol to do so that uses $\Oish(n)$ bits of communication per source node, based on a slight tweak of a protocol described in \cite{FWY20} (c.f.~Lemma 9).
Thus $|V'| \cdot \Oish(n) = \Oish(n^{5/3})$ bits are communicated in total.

At the end of the protocol, we claim that both Alice and Bob have sufficient information to each compute a state of the art shortcut set for the original graph in a black-box way.
More specifically, letting $G' \subseteq G$ be the subgraph containing only the edges in the BFS trees that they have computed, we claim that (with high probability) a $O(n^{1/3})$-diameter-reducing shortcut set for $G'$ is also an $O(n^{1/3})$-diameter-reducing shortcut set for $G$.
To see that this is the case, notice that for every pair of nodes $u,v$ in $G$, if we sample a node $x \in V'$ that lies along a $u \leadsto v$ path, then there will still be a $u \leadsto v$ path in $G'$, and so the shortcut set for $G'$ will reduce $\dist(u, v)$ to at most $O(n^{1/3})$. 
Since our sampling rate is $|V'| = \Oish(n^{2/3})$, with high probability, we will sample such a node $x$ for all pairs with $\dist_G(u, v) \ge n^{1/3}$.

\subsubsection{Key Ideas for a $\Oish(n^{4/3})$ Communication Bound}



Now we will describe the weakness in the straightforward protocol that we plan to improve.
This protocol does not take advantage of the technical details of the current state-of-the-art centralized shortcut set construction by Kogan and Parter \cite{KP22a} (henceforth, the \emph{KP algorithm}).
We will apply the KP algorithm to the communication setting in a white-box manner.

The natural implementation of the KP algorithm still turns out to be communication-heavy.
The main communication bottleneck turns out to be the computation of a \emph{path cover}, a basic technical ingredient used in the KP algorithm.
A path cover $Q$ is a set of paths in the transitive closure $G^*$ with the following properties:
\begin{itemize}
\item The paths in $Q$ are pairwise node-disjoint,
\item $Q$ has size $|Q| \le O(n^{2/3})$,
\item For each path $\pi$ in $G$, at most $O(n^{1/3})$ nodes in $\pi$ are not contained in any of the paths $q \in Q$.
\end{itemize}
A direct implementation of the folklore greedy path cover algorithm requires Alice and Bob to find two reachable nodes at distance $\Omega(n^{1/3})$ for each path in the cover.
The most natural way to do so involves solving a single-source reachability problem for each path.
Using the $\Oish(n)$-bit single-source protocol described above separately for each of the $O(n^{2/3})$ paths gives a communication bound of $\Oish(n^{5/3})$, matching the straightforward algorithm sketched previously.
In the centralized setting, faster path-cover constructions are known, including the min-cost-flow-based approach of Kogan and Parter~\cite{kogan_beating_2022}. Their running-time guarantees do not by themselves yield low-communication implementations.

Our main technical contribution is a new dynamic data structure that allows Alice and Bob to compute a path cover with only $\Oish(n^{4/3})$ bits of communication.
Our key idea is to compute the path cover on the basis of \emph{longest} paths between node pairs, rather than \emph{shortest} paths.
The advantage of doing so is that updating the data structure after every new path is added to the cover would be expensive if we focused on shortest paths, but the updates are cheaper when focusing on longest paths.

The bulk of the cost when focusing on shortest paths comes from the fact that longest shortest path distance calculations from a specific node can be unrelated to longest shortest path distance calculations from its reachable neighbors. 
Therefore, naively, it seems necessary to compute a separate reachability tree for each node.

On the other hand, the maximum length of any path starting from each node can be computed by only focusing on the immediate out-neighbors of the node. 
This is particularly good for the communication complexity setting, as the players can perform unlimited computation locally based on their own edges and limit communication only to when a change to the data structure is necessary. 





In addition to the new data structure, we use one more technical departure from the KP algorithm.
After computing the path cover, the KP algorithm includes a shortcutting step that requires computing reachability between specific pairs of nodes.
The KP algorithm adds shortcut edges from a random sample of $|V'| = \Oish(n^{2/3})$ nodes to certain nodes along a random sample of $|Q'| = O(n^{1/3})$ paths from the cover; doing so requires us to fully compute the reachability relation between nodes in $V'$ and nodes that lie on paths in $Q'$.
A naive implementation performs a BFS search from each $v \in V'$, or a reverse-BFS search from each node on a path in $Q'$, but both of these strategies would again require $\Oish(n^{5/3})$ bits of communication.
We improve this to $\Oish(n^{4/3})$ bits using a \emph{path-BFS} procedure, which is run from each path $q \in Q'$ rather than each node $v \in V'$.
Details are provided in Section \ref{sec:pathBFS}.
Related centralized computations of the first reachable vertex on a path appear in Kogan and Parter~\cite[Lemma~3.1]{kogan_beating_2022}.

\subsection{Lower Bound Approach \label{sec:lboverview}}

We obtain our communication lower bound by employing path systems with useful structure that were shown to recover the $\Omega(n^{1/6})$ lower bound for shortcut sets in the centralized setting \cite{BHT23}. More specifically, we embed one large set-disjointness instance into the edges of an acyclic,
$\infty$-bridge-free path system. 
The key property is that every subpath of a system path is the unique directed path between its
endpoints in the union graph. 
Consequently, a valid shortcut whose endpoints lie on a system path certifies that every edge of the intervening subpath is present in the input graph.

\begin{definition}
    In $\mathrm{DISJ}_m$, Alice and Bob receive vectors $x,y\in\{0,1\}^m$ and must decide whether there is no coordinate $i$ with $x_i=y_i=1$. For any fixed error probability below $1/2$, this requires $\Omega(m)$ bits of communication.
\end{definition}

The above problem is widely studied, and has found various applications (see for example \cite{kushilevitz1997communication,kalyanasundaram1992probabilistic,razborov1992distributional,bar2004information,chattopadhyay2010set,weinstein2015simultaneous}).

Encode $\mathrm{DISJ}_m$ on the $m$ edges of a public path system
with $p$ paths: Alice includes $e$ iff $x_e=0$, and Bob iff
$y_e=0$. Missing edges thus correspond to intersecting coordinates.
By uniqueness, a correct $D$-diameter-reducing shortcut set certifies the presence of every edge it bypasses along a system path.

While the most current shortcut set lower bounds do not use bridge-free path systems (see \cite{BH23}), the bridge-free path system constructions are better suited for showing stronger communication complexity lower bounds because of their size. To our knowledge, this is the first time that bridge-free path systems are used to establish communication complexity lower bounds, and it is an interesting question whether they might be applicable to more problems in the area. 

\section{Preliminaries}
The graphs in this paper are directed. We sometimes use the notation $G = G_A \cup G_B$. Here $G_A := (V, E_A)$, $G_B := (V, E_B)$, and $G_A \cup G_B = (V, E_A \cup E_B)$.

The concept of communication complexity was first introduced by Yao \cite{Yao1979CommunicationComplexity}. We use the standard two-party communication model with shared
randomness; see~\cite{kushilevitz1997communication}. Alice and Bob receive private inputs,
perform unrestricted local computation, and communicate by exchanging
bits. The communication cost is the maximum number of bits exchanged
over all inputs and all outcomes of the randomness.

For the shortcut-set problem, the protocol succeeds when both players
output the same valid shortcut set. A protocol has error at most
$\rho$ if it succeeds with probability at least $1-\rho$ on every
input. For a communication problem $\mathcal P$, we write
$R_\rho(\mathcal P)$ for the minimum communication cost of a protocol
solving $\mathcal P$ with error at most $\rho$.

\section{Shortcut Set Upper Bound: $D \leq n^{1/3}$}
To prove Theorem \ref{thm:main-ub}, we show that it suffices to focus on the case where the input graph is acyclic and the length of any path in the graph is bounded. This is proved in Section \ref{sec:reds}. More concretely, for the rest of this section, we assume that the input graph $G = G_A \cup G_B$ is 
\begin{enumerate}
    \item acyclic and
    \item for any directed path $\pi$ in $G$, the number of nodes in $\pi$ is at most $O(n/D)$, where $D$ is the target diameter bound.
\end{enumerate}

We will consider the following computational problems in the 2-party communication setting and use communication bounds on these problems to prove Theorem \ref{thm:main-ub}.

\begin{definition}[\textsc{PathCover}]
\label{def:PathCover} 
In the $\textsc{PathCover}$ computational problem in the 2-party communication setting, two players (Alice and Bob) are given a parameter $D$ and directed edge sets $E_A$ and $E_B$ on $n$ nodes $V$. The goal is to compute a set of paths $Q$ in the transitive closure of $G = (V, E = E_A \cup E_B)$, so that 
\begin{enumerate}[(1)]
    \item the paths in $Q$ are node disjoint,
    \item for every path $\pi$ in $G$ with $|\pi| \geq D$, at most $D/2$ nodes are not in $V(Q)$, and
    \item for every path $q \in Q$, the length of the path is $|q| \geq D/2$.
\end{enumerate}
\end{definition}

The above concept is related to chain-antichain decompositions appearing in previous work on shortcut sets (for example, \cite{KP22a}).

\begin{definition} [\textsc{PathBFS}]
\label{def:PathBFS}
In the $\textsc{PathBFS}$ computational problem in the 2-party communication setting, two players (Alice and Bob) are given directed edge sets $E_A$ and $E_B$ on $n$ nodes $V$, and a directed path $\pi$. The edges along $\pi$ may or may not be in $E_A \cup E_B$, but for each consecutive pair of vertices $x_1, x_2$ along $\pi$, it is guaranteed that there is an $x_1$ to $x_2$ path in $G  = (V, E = E_A \cup E_B)$. The goal is to compute a subgraph $T$ of $\pi \cup G$ with the following property: 
\begin{itemize}
    \item For every pair of vertices $(u,v) \in \pi \times V$ (or $\in V \times \pi$), if there is a $u$ to $v$ path in $G$, then there is also a $u$ to $v$ path in $T$. 
\end{itemize}
\end{definition}

We describe the algorithm for $8\le D\le n^{1/3}$;
smaller values of $D$ are handled by the constant-diameter
case in Section~\ref{sec:reds}.
Fix a sufficiently large absolute constant $c$.
All random choices are public, and the vertex sample is
independent of the path sample.
For a path $\pi$, let $E(\pi)$ denote its consecutive-vertex
edges.







    

    

\begin{algorithm}[H]
    \DontPrintSemicolon
    \SetKwInOut{Input}{Input}
    \SetKwInOut{Output}{Output}

    \caption{2-Player Shortcut Set Computation
    \label{alg:main-alg}}

    \Input{An $n$-vertex DAG $G=G_A\cup G_B$ and
    $8\le D\le n^{1/3}$}
    \Output{An edge set $H\subseteq G^*$ that reduces
    the hop diameter to at most $D$ with high probability}

    $\mathcal{Q}\gets\texttt{PathCover}(G,D)$\;

    \ForEach{$\pi\in\mathcal{Q}$}{
        Construct a canonical two-hop shortcut set
        $S_\pi$ for $\pi$, with $O(|V(\pi)|\log n)$
        edges, as in~\cite{Raskhodnikova10}\;
    }

    $H_0\gets
    \bigcup_{\pi\in\mathcal{Q}}(E(\pi)\cup S_\pi)$\;
    $H\gets H_0$\;

    $p\gets\min\{1,c\log n/D\}$\;
    Form $\mathcal{Q}'$ by including $\lceil p \cdot |\mathcal{Q}| \rceil$ paths from $\mathcal{Q}$ uniformly at random\;
    \ForEach{$\pi\in\mathcal{Q}'$}{
        $T_\pi\gets\texttt{PathBFS}(\pi,G)$\;
    }

    $G'\gets
    \left(V,\bigcup_{\pi\in\mathcal{Q}'}E(T_\pi)\right)$\;

    Choose $V'$ by including $\lceil pn \rceil$ vertices from $V$ uniformly at random\;

    \ForEach{$v\in V'$ and $\pi\in\mathcal{Q}'$}{
        \If{some vertex of $\pi$ is reachable from $v$
        in $G'$}{
            Let $u$ be the first such vertex along $\pi$\;
            \If{$u\ne v$}{
                $H\gets H\cup\{(v,u)\}$\;
            }
        }
    }

    \Return{$H$}\;
\end{algorithm}

We note that Algorithm \ref{alg:main-alg} makes calls to \texttt{PathBFS} and \texttt{PathCover}, which are interactive protocols between Alice and Bob that are defined in Sections \ref{sec:pathBFS} and \ref{sec:pathcover}. These protocols solve the computational problems with the same names, as specified in Definitions \ref{def:PathBFS} and \ref{def:PathCover}.

\begin{lemma}
Algorithm \ref{alg:main-alg} outputs a $D$-diameter reducing shortcut set with high probability.
\end{lemma}
\begin{proof}
    We note that the underlying shortcut set algorithm in this communication protocol is the algorithm from \cite{KP22a}, which we recap in \ref{sec:ubapproach}, and so the proof of correctness from \cite{KP22a} holds without significant modification.
    However, we remark that there are two minor deviations.
    One is that \cite{KP22a} uses a slightly different parametrization ($D/8$ rather than $D$).
    The other is that, to check connectivity between the sampled nodes $V'$ and the sampled paths $\mathcal{Q}'$, the two players compute a subgraph $G'$ of $G^*$. Due to the properties of $\texttt{PathBFS}$ stated in Definition \ref{def:PathBFS}, we can conclude that for every vertex $v \in V$ and $u \in V(\mathcal{Q}')$, if there is a $v$ to $u$ path in $G$ then there is also a $v$ to $u$ path in $G'$. Therefore, the proof of correctness given in \cite{KP22a} applies without significant modification.
\end{proof}

\begin{lemma}
    \label{lem:CCSS-to-CCPC}
    Algorithm \ref{alg:main-alg} computes an edge set $H$ of size
        $$|H| \leq \Oish \left ( \frac{n^2}{D^3} \right)$$
    using 
        $$\Oish \left( \frac{n}{D^2} \right) \cdot \texttt{CC}(\textsc{PathBFS}) + \texttt{CC}(\textsc{PathCover})$$
    bits of communication (deterministically).
\end{lemma}

\begin{proof}
    There are at most
    $$|H_0| + |V'| \cdot |\mathcal{Q}'| \leq \Oish(n) + \Oish(n^2/D^3)$$
    edges in $H$; note that $D \le n^{1/3}$, so the latter term dominates.
    
    Note that there are two steps that require any communication: the calls to \texttt{PathCover} and \texttt{PathBFS}. The cost of the call to \texttt{PathCover} is bounded by \texttt{CC}(\textsc{PathCover}). There are at most $\Oish(n/D^2)$ calls to \texttt{PathBFS}, so their cost is bounded by $\Oish(n/D^2) \cdot$ \texttt{CC}(\textsc{PathBFS}).
\end{proof}

\begin{lemma}
    \label{lem:CCPC}
    $\texttt{CC}(\textsc{PathCover}) \leq  \Oish \left( nD + n^2/D^2  \right)$
\end{lemma}

\begin{lemma}
    \label{lem:CCPathBFS}
    $\texttt{CC}(\textsc{PathBFS}) \leq  \Oish \left( n \right)$
\end{lemma}

We defer the proofs of Lemmas~\ref{lem:CCPC} and~\ref{lem:CCPathBFS} to Sections~\ref{sec:pathcover} and~\ref{sec:pathBFS}, respectively. For now, we show how they imply the main upper bound result in our paper.

\begin{proof}[\textit{(Proof of Theorem \ref{thm:main-ub} using Lemmas \ref{lem:CCPC} and \ref{lem:CCPathBFS}).}]
    Plugging in the inequalities from Lemmas \ref{lem:CCPC} and \ref{lem:CCPathBFS} into the bit upper bound from Lemma \ref{lem:CCSS-to-CCPC}, we conclude that, given any graph $G$ and a parameter $D \leq n^{1/3}$ in the 2-player communication complexity setting, there is a 2-player communication protocol that uses 
    $$\Oish \left( \frac{n^2}{D^2} \right) + \Oish(nD) \leq \Oish \left( \frac{n^2}{D^2} \right)$$
    bits and returns a set of edges $H$ with size $|H| \leq \Oish(n^2/D^3)$ that is a $D$-diameter reducing shortcut set for $G$ with high probability. 
\end{proof}

\subsection{Implementation of PathBFS \label{sec:pathBFS}}
Our implementation of \texttt{PathBFS} requires a communication protocol to solve the usual single node BFS problem in the 2 player communication complexity setting. The following definition specifies the $\textsc{BFS}_{out}$ problem. The definition of $\textsc{BFS}_{in}$ is analogous.

\begin{definition}[$\textsc{BFS}_{out}$] \label{def:standardBFS}
    In the $\textsc{BFS}_{out}$ problem in the 2-party communication setting, two players (Alice and Bob) are given directed edge sets $E_A$ and $E_B$ on $n$ nodes $V$, and a node $u \in V$. The goal is to compute a subgraph $T$ of $G = (V, E_A \cup E_B)$ so that for each pair of nodes $(u,v)$, where $u$ is the input node and $v$ is any node in $V$, it is the case that $$\dist_{T}(u,v) = \dist_G(u,v).$$
\end{definition}

\begin{lemma}[\cite{FWY20}] \label{lem:standardBFS}
    Given an instance of $\textsc{BFS}_{in/out}$, there exists a 2 player communication protocol that outputs a solution $T$ to the problem using $\Oish(|T|)$ bits.
\end{lemma}

For the implementation of a protocol that solves $\textsc{BFS}_{out/in}$ using $\Oish(|T|)$ bits of communication please refer to Lemma 9 and its proof in \cite{FWY20}. Their algorithm is tailored to a slightly different setting, but it can also solve the problem in Definition \ref{def:standardBFS} without significant edits (see their Appendix A).  

\begin{algorithm}[H]
    \DontPrintSemicolon
    \SetKwInOut{Input}{Input}
    \SetKwInOut{Output}{Output}

    \caption{\texttt{PathBFS} Interactive Protocol \label{alg:pathBFS}}

    \Input{Directed graph $G = G_A \cup G_B$, directed path $\pi = (x_1, \dots, x_{|\pi|})$ in $G^*$}
    \Output{A subgraph $T\subseteq G\cup\pi$ satisfying Definition~\ref{def:PathBFS}}

    ~\\

    $T_{in}, T_{out} \gets \emptyset$\;

    ~\\
    
    \ForEach{$i \in \{1, \dots, |\pi|\}$ in increasing order}{
            $T_{in} \gets T_{in} \cup \texttt{BFS}_{in}(x_i, \ G \setminus V(T_{in}))$
        }

    ~\\
    
    \ForEach{$i \in \{1, \dots, |\pi|\}$ in decreasing order}{
            $T_{out} \gets T_{out} \cup \texttt{BFS}_{out}(x_i, \ G \setminus V(T_{out}))$
        }

    ~\\
    
    $T \gets \pi \cup T_{in} \cup T_{out}$
\end{algorithm}

\begin{lemma} \label{lem:pathBFS-alg}
Algorithm \ref{alg:pathBFS} correctly solves \textsc{PathBFS}, and uses $\Oish(n)$ bits of communication.
\end{lemma}

\begin{proof}
    To see that the algorithm correctly solves $\textsc{PathBFS}$, note that for each node $u \in \pi$, Alice and Bob run a $\texttt{BFS}_{out}$ interactive protocol rooted at $u$ at some point during the algorithm. Therefore, if $u$ is the last node along $\pi$, then for each pair of vertices $(u,v) \in \pi \times V$ with a $u \rightsquigarrow v$ path in $G$, some $u$ to $v$ path was added to $T_{out}$ by the correctness of $\texttt{BFS}_{out}$. If $u$ is not the last node along $\pi$, let $x_1, x_2, \dots$ be the nodes that come after $u$ along $\pi$. If there is a $u \rightsquigarrow v$ path in $\texttt{BFS}_{out}(u)$, then we are done. Otherwise, there exists a vertex $x_i$ that comes after $u$ along $\pi$ such that there is a $x_i \rightsquigarrow v$ path in $\texttt{BFS}_{out}(x_i)$. Therefore, there is a $u \rightsquigarrow x_i \rightsquigarrow v$ path in $T_{out} \cup \pi \subseteq T$. The analogous is true for each pair of vertices $(u,v) \in V \times \pi$. \vskip 10pt
    
    Within the $\texttt{BFS}_{out}$ pass, each search is run after deleting all vertices in the outputs of earlier outgoing searches.
    Thus the output vertex sets are pairwise disjoint, and their total size is at most $n$. 
    The same argument applies separately to the $\texttt{BFS}_{in}$ pass. Although incoming and outgoing searches may overlap, the total number of vertices summed over all search outputs is therefore at most $2n$.
    By Lemma~\ref{lem:standardBFS}, the total communication is $\Oish(n)$ bits.
\end{proof}

\subsection{Implementation of PathCover using a Data Structure \label{sec:pathcover}}

In this section, we specify a data structure jointly maintained by two players in the 2-party communication complexity setting that supports a vertex-marking operation using bounded communication for updates. Assuming the existence of this data structure, we give an algorithm that queries this data structure to compute a path cover of a given directed acyclic graph $G$ using bounded communication. In the next section, we will prove that this data structure actually exists.

\begin{lemma} \label{lem:datastructure}
    In the 2-player CC model, there is a data structure $\mathcal{D}$ shared between the two players (say Alice and Bob) that preprocesses an $n$-node graph $G = G_A \cup G_B$ with a fixed parameter $D \leq n^{1/3}$, so that after preprocessing 
    \begin{enumerate}[(1)]
        \item All nodes are initially unmarked.
        \item The data structure supports operation QueryPath($\mathcal{D}$) that does one of the following
        \begin{enumerate}[(a)]
            \item Return a path $\pi$ in $G$ that contains at least $D/2$ unmarked nodes and marks all the unmarked nodes along $\pi$.
            \item Reports that there is no such path.
        \end{enumerate}
        \item The total bits of communication across any sequence of operations is 
        $$\Oish \left( nD + \frac{n^2}{D^2} \right)$$
    \end{enumerate}
\end{lemma}

Assuming the existence of the data-structure in Lemma \ref{lem:datastructure}, Alice and Bob can compute a path cover for $G = G_A \cup G_B$ as described in Algorithm \ref{alg:pathcover}. In the algorithm, for each $\pi$ returned by QueryPath($\mathcal{D}$), we use $\texttt{new}(\pi)$ to refer to the nodes along $\pi$ that were newly marked by the corresponding QueryPath($\mathcal{D}$) operation.


\begin{algorithm}[H] \label{alg:pathcover}
    \DontPrintSemicolon
    \SetKwInOut{Input}{Input}
    \SetKwInOut{Output}{Output}

    \caption{PathCover($G$, $D$)}

    \Input{Directed acyclic graph $G = G_A \cup G_B$, parameter $D$}
    \Output{Path cover $\mathcal{Q}$}

    \vskip10pt
    Initialize a data structure $\mathcal{D}$ with parameter $D$ from Lemma \ref{lem:datastructure}\;
    $\mathcal{Q} \gets \emptyset$\;

    \vskip10pt
    $\pi \gets $ QueryPath($\mathcal{D}$)\;
    \While{$\pi \ne $ NULL}{
        add $\texttt{new}(\pi)$ to $\mathcal{Q}$\;
        $\pi \gets $ QueryPath($\mathcal{D}$)\;
    } 

    \Return{$\mathcal{Q}$}
\end{algorithm}

\begin{lemma} \label{lem:pathcovcorrectness}
    Given a DAG $G$ in the 2-party CC setting, Algorithm \ref{alg:pathcover} computes a correct path cover of $G$ using $$\Oish \left(nD + \frac{n^2}{D^2} \right)$$ bits of communication. 
\end{lemma}

\begin{proof}[(Proof of Lemma \ref{lem:pathcovcorrectness} assuming Lemma \ref{lem:datastructure})]
    We first note that a set of paths $\mathcal{Q}$ returned by Algorithm \ref{alg:pathcover} is a solution to the $\textsc{PathCover}$ problem defined in Definition \ref{def:PathCover}. Note that the paths in $\mathcal{Q}$ meet requirements (1) and (3) of Definition \ref{def:PathCover} by construction of QueryPath($\mathcal{D}$). Since the algorithm continues until there are no more paths in $G^*$ with $\geq D/2$ unmarked nodes, property (2) of Definition \ref{def:PathCover} is also satisfied. Moreover, the only steps of the algorithm that require communication between Alice and Bob are the calls to QueryPath($\mathcal{D}$). Thus, by property (3) of Lemma \ref{lem:datastructure}, the total amount of communication is upper-bounded by $\Oish(nD + n^2/D^2)$.
\end{proof}

\subsection{Implementation of Data Structure in Lemma \ref{lem:datastructure} \label{sec:data-str-imp}}

\subsubsection{List of Data} To implement a data structure $\mathcal{D}$ that meets the criteria in Lemma \ref{lem:datastructure}, Alice and Bob will each maintain the following data:
\begin{itemize}
    \item For each node $v$, Alice and Bob maintain a bool $\texttt{marked}(v)$, which signifies whether $v$ is marked.

    \item Let us say that a path is \emph{$E_A$-starting} (or \emph{$E_B$-starting}) if it consists of a single vertex or its first edge belongs to $E_A$ ($E_B$).    
    For each node $v \in V$, Alice and Bob will maintain integer labels $\ell_A(v)$ and $\ell_B(v)$, each in the range $[0, D]$.
    The label $\ell_A(v)$ will be the minimum of $D$, and the largest number of unmarked nodes in any $E_A$-starting path whose first node is $v$.
    The label $\ell_B(v)$ is defined similarly (with $E_B$-starting paths).
    We will also write $\ell(v) := \max\{\ell_A(v), \ell_B(v)\}$.
\end{itemize}

\subsubsection{Preprocessing} 
In the preprocessing step, Alice and Bob aim to initialize a data structure $\mathcal{D}$ from Lemma \ref{lem:datastructure} while limiting the amount of communication exchanged.

\begin{lemma} \label{lem:ds-preprocessing}
    For any directed, acyclic $n$-node graph $G$ and parameter $D \leq n^{1/3}$ in the 2-party communication complexity model, there is an $\Oish(nD)$ bit protocol to set up a data structure $\mathcal{D}$ from Lemma \ref{lem:datastructure}.
\end{lemma}

\begin{proof}
    Consider the following protocol:
    \begin{itemize}
        \item All nodes begin unmarked.
        \item For each vertex $v$, initialize its labels $\ell_A(v)$ and $\ell_B(v)$ to $1$.
    
        \item Alice and Bob alternate sending messages as follows.
        On Alice's turn, for every node $v$ with $N_A^{out}(v) \neq \emptyset$, she computes
        $$\ell^{new}_A(v) = \min \left\{ D, \ \ 1 + \max_{u \in N^{out}_A(v)} \ell(u) \right\}.$$ 
        Then, she sends over the names and new labels for any nodes $v$ with $\ell^{new}_A(v) \ne \ell_A(v)$.
        Alice and Bob both update $\ell_A(v) \gets \ell_A^{new}(v)$ for any such nodes.
        
        \item Bob's message is analogous.

        \item The process halts once Alice and Bob send over an empty list on consecutive rounds.
    \end{itemize}

    Note that each node $v$ will be sent at most $O(D)$ times, since its label $\ell_A(v)$ or $\ell_B(v)$ increases each time it is sent, but cannot exceed $D$. The number of bits communicated across all empty list messages is also $\Oish(nD)$ (note that we consider an empty list to be a special message, which requires $\Theta(1)$ bits to distinguish it from other messages). The upper bound holds because for each Alice-Bob pair of messages sent, at least one node's label increases by $\geq 1$. \vskip 10pt

    We next argue correctness of the labels set by the above protocol:

    \begin{claim} \label{clm:label-correctness-preproc}
        For all integers $k \in [1, D]$ and all nodes $v$, the label of $v$ is $\ell(v) \geq k$ iff there is a path in $G$ that starts with $v$ and uses $k$ unmarked nodes.
    \end{claim}

    \begin{claimproof} ~
    \begin{itemize}
    \item ($\rightarrow$) Suppose $v$ has label $\ell(v) \geq k$. We would like to show that there exists a path in $G$ that starts with $v$ and has $k$ unmarked nodes. In the case that $k = 1$, for any node $v$ with $\ell(v) = 1$, the one-node path $\pi = (v)$ has exactly $1$ unmarked node. Then, we consider $k > 1$. For any node $v$ with $\ell(v) \geq k$, there is a node $u$ in its out-neighborhood that has label $\ell(u) \geq k-1$. Applying the inductive hypothesis, let $\pi$ be a path in $G$ that starts with $u$ and has $\geq k-1$ unmarked nodes. Concatenating $v$ to the start of $\pi$ gives a path in $G$ that starts with $v$ and has $\geq k$ unmarked nodes.

    \item ($\leftarrow$) Suppose that for a node $v$ and an integer $k \in [1, D]$, there exists a path in $G$ that starts with $v$ and has $k$ unmarked nodes. The claim can be proved by induction. In the case that $k = 1$, for every vertex $v$, its label $\ell(v)$ is automatically $\geq 1$, as $1$ is the starting value for every vertex. We now focus on the case $k > 1$. For any node $v$, assume there exists a path in $G$ that starts with $v$ and has $k$ unmarked nodes. Then there exists an out-neighbor $u$ of $v$, so that there is a path in $G$ that starts at $u$ and uses $k - 1$ unmarked nodes. By inductive hypothesis, $u$ has label $\ell(u) \geq k - 1$. Therefore, according to the label setting protocol, either $\ell_A(v) \geq k$ or $\ell_B(v) \geq k$. \qedhere
\end{itemize}
\end{claimproof}
\end{proof}

\subsubsection{Marking Nodes}

After marking (possibly batches of) nodes, the following lemma summarizes how Alice and Bob maintain the data structure: 

\begin{lemma} \label{lem:ds-marking}
    Given a data structure $\mathcal{D}$ specified in Lemma \ref{lem:datastructure}, there exists a 2-party communication protocol that guarantees that the properties specified in Lemma \ref{lem:datastructure} continue to hold for $\mathcal{D}$ after every node-marking operation, and which uses $\Oish(nD)$ bits of communication across any sequence of node-marking operations.
\end{lemma}

\begin{proof}
    Consider the following protocol between Alice and Bob:

    \begin{itemize}
        \item For a vertex $v$ that was marked, set its bool $\texttt{marked}(v)$ to true.
        \item Alice and Bob exchange messages until they both send empty lists in two consecutive messages. Each message exchange proceeds as follows:
        \begin{itemize} 
            \item When it is Alice's turn to send a message, for every node $v$, she computes $\ell_A^{\mathrm{new}}(v)$ as follows: 
            \begin{enumerate}
                \item If $\texttt{marked}(v) = false$, then 
                $$\ell^{new}_A(v) = 
                \begin{cases}
                    \ell_A(v) &  \text{if }N_A^{out}(v) = \emptyset \\
                    \min \{ \ell_A(v), \ \ 1 + \max_{u \in N^{out}_A(v)} \ell(u) \} & \text{otherwise.}
                \end{cases}$$
                \item If $\texttt{marked}(v) = true$, then 
                $$\ell^{new}_A(v) = 
                \begin{cases}
                    0 &  \text{if }N_A^{out}(v) = \emptyset \\
                    \min \{ \ell_A(v), \ \ \max_{u \in N^{out}_A(v)} \ell(u) \} & \text{otherwise}.
                \end{cases}
                $$
            \end{enumerate}
            Then, in the message, she sends the node names and new labels for any nodes $v$ with $\ell^{new}_A(v) \ne \ell_A(v)$. If the list of nodes sent by Alice is not empty, then for all such nodes $v$, Alice and Bob update $\ell_A(v) \gets \ell_A^{new}(v)$.
            \item Bob's message is analogous.
        \end{itemize}
    \end{itemize}

    First, note that for each node $v$, its labels $\ell_A(v) \in [0, D]$ and $\ell_B(v) \in [0, D]$ can only decrease. Therefore, across all node-marking operations, a node $v$ can be sent in at most $\Oish(D)$ messages. Additionally, the number of bits required across all empty list messages is at most $\Oish(nD)$, since for every pair of consecutive messages following a node-marking operation (except the last pair), at least one of the messages is a node with a newly updated label.

    To show correctness of the new labels after each node-marking operation, we prove the following claim:

    \begin{claim} \label{clm:label-correctness-updates}
        For all integers $k \in [0, D]$ and all nodes $v$, the label of $v$ is $\ell(v) \geq k$ iff there is a path in $G$ that starts with $v$ and uses at least $k$ unmarked nodes.
    \end{claim}

    \begin{claimproof}~   
    After the update protocol terminates, we prove the claim for all $k\in[0,D]$ at once, starting with vertices of out-degree zero and then considering each vertex after the claim has been proved for all its out-neighbors. This is possible because $G$ is acyclic.
        \begin{itemize}
            \item ($\leftarrow$) For any $k \in [0, D]$ and node $v$, suppose that there exists a path in $G$ that starts with $v$ and has at least $k$ unmarked nodes. In the case that $k = 0$, it trivially holds that for every node $v$, its label is $\ell(v) \geq 0$. We now discuss the case that $k > 0$. For any node $v$, assume there exists a path in $G$ that starts with $v$ and has at least $k$ unmarked nodes. Then, there are two cases. One case is that $N^{out}(v) = \emptyset$. Then, it must be true that $v$ is unmarked and $\ell(v) = k = 1$. The second case is that $N^{out}(v) \neq \emptyset$. Then, since $k \geq 1$, there exists an out-neighbor $u$ of $v$, such that there's a path in $G$ that starts with $u$ and uses $\geq k - 1$ unmarked nodes if $v$ is unmarked and $\geq k$ unmarked nodes if $v$ is marked. By the inductive hypothesis, $u$ has label $\ell(u) \geq k - 1$ if $v$ is unmarked or $\ell(u) \geq k$ if $v$ is marked. Therefore, according to the label update protocol, either $\ell_A(v) \geq k$ or $\ell_B(v) \geq k$.
            \item ($\rightarrow$) In the other direction, suppose $v$ has label $\ell(v) \geq k$. We would like to show that there exists a path in $G$ that starts with $v$ and has at least $k$ unmarked nodes. In the case that $k = 0$, it trivially holds that any path starting with $v$ has $\geq 0$ unmarked nodes. In the case that $k > 0$, there are two possibilities. The first is that $N^{out}(v) = \emptyset$. Then, $v$ is unmarked and $k = 1$. The one-node path $\pi = (v)$ confirms the claim. The second possibility is that $N^{out}(v) \neq \emptyset$. Since $N^{out}(v) \neq \emptyset$, note that for a node $v$ with $\ell(v) \geq k$, there is a node $u$ in its out-neighborhood that has label $\ell(u) \geq k-1$ if $v$ is unmarked or $\ell(u) \geq k$ if $v$ is marked. Let $\pi$ be a path in $G$ that starts with $u$ and has $ \geq k-1$ or $\geq k$ unmarked nodes, in the cases that $v$ is unmarked and marked respectively. Concatenating $v$ to the start of $\pi$ gives a path in $G$ that starts with $v$ and has $\geq k$ unmarked nodes. \qedhere
        \end{itemize}
    \end{claimproof}
\end{proof}

\subsubsection{Implementing QueryPath($\mathcal{D}$)}
Recall that the QueryPath($\mathcal{D}$) operation consists of finding a path with $\geq D/2$ uncovered nodes, followed by marking all vertices along the path. So far, we gave the two-party communication protocol used to maintain the data structure $\mathcal{D}$ after vertex-marking operations. It remains to show a communication-efficient protocol that finds a path in $G^*$ with $\geq D/2$ uncovered nodes.

To return a path with at least $D/2$ unmarked nodes, Alice and Bob can start by finding a vertex $v$ with a maximum label among all the vertices. There are two cases, depending on the label of $v$:
\begin{itemize}
    \item Case 1 is that $\ell(v) < D/2$. For this case, we will show that there is no path in $G$ containing $\geq D/2$ unmarked nodes.
    \item Case 2 is that $\ell(v) \geq D/2$. For this case, we will give a two-party protocol to find a path with $\geq D/2$ unmarked nodes that starts with $v$.
\end{itemize}

\begin{lemma} \label{lem:no-long-unmarked-path}
    If at any point every node $v$ has label $\ell(v) < D/2$, then there is no path in $G$ with $\geq D/2$ unmarked nodes.
\end{lemma}

\begin{proof}
    This follows directly from Claims \ref{clm:label-correctness-preproc} and \ref{clm:label-correctness-updates}. The invariants on label values imply that the value of its label $\ell(v)$ is equal to the maximum number of unmarked nodes that can be found along a simple path in $G$ that has $v$ as its first node.
\end{proof}

The rest of this section focuses on giving the details of the path finding process in the case that there exists a node $v$ with label $\ell(v) \geq D/2$.

\begin{lemma} \label{lem:find-long-path}
    For every integer $k \in [0, n]$, if there exists a node $v$ with label $\ell(v) \geq k$, then there is a two-party communication protocol that finds a path $\pi$ with $\geq k$ unmarked nodes using $\Oish(|\pi|)$ bits of communication.
\end{lemma}

\begin{proof}
    For a fixed integer $k$, if there exists a node $v$ with label $\ell(v) \geq k$, then Alice and Bob can find a path $\pi$ in $G$ with $\geq k$ unmarked nodes using the following protocol:
    \begin{itemize}
        \item Let the first node along $\pi$ be $v$. Alice and Bob will repeatedly exchange messages to pick the next node to add to $\pi$ until the last node along $\pi$ has out-degree $0$.
        \item Let $u$ be the last node along $\pi$. Suppose $u$ has out-degree $>0$ in $G$.
        \item Both Alice and Bob will pick and send a vertex $u^{next}_A \in N^{out}_A(u)$ and $u^{next}_B \in N^{out}_B(u)$ with 
        $$\ell(u_A^{next}) \geq \max_{y \in N^{out}_A(u)} \ell(y) \ \ \text{ and } \ \ \ell(u_B^{next}) \geq \max_{y \in N^{out}_B(u)} \ell(y).$$ 
        Or send $NULL$ if $u$ has out-degree zero in their respective subgraph.
        \item The next node along $\pi$ is whichever node among $u_A^{next}$ and $u_B^{next}$ has the higher label.
    \end{itemize}

    The protocol uses $\Oish(|\pi|)$ bits of communication because for every round of message exchanges, at least one new node is concatenated to the end of $\pi$.
    
    To prove that the final path $\pi$ contains at least $k$ unmarked nodes, observe that if a node $u$ on $\pi$ is marked, then its successor on $\pi$ has label $\ell(u)$, whereas if $u$ is unmarked, then its successor has label at least $\ell(u)-1$. It follows that along $\pi$, the label can either remain unchanged or decrease by exactly $1$, and the latter can occur only at an unmarked node. Let (z) be the last node of $\pi$. There are exactly $\ell(v)-\ell(z)$ label decreases along $\pi$, each accounting for a distinct unmarked node before $z$. Since $z$ has out-degree zero, $\ell(z)=1$ if $z$ is unmarked and $\ell(z)=0$ otherwise. Counting $z$ as well, the total number of unmarked nodes is therefore at least
    $$\ell(v)-\ell(z)+\ell(z)=\ell(v)\ge k.$$
\end{proof}

\begin{lemma} \label{lem:total-path-finding}
    Across any sequence of QueryPath($\mathcal{D}$) operations, Alice and Bob spend $\Oish \left( n^2/D^2 \right)$ bits of communication on finding all of the paths returned in this sequence (not including the updates to the data structure from vertex-marking operations).
\end{lemma}

\begin{proof}
    Each QueryPath($\mathcal{D}$) operation finds and marks at least $D/2$ nodes, so there are at most $O(n/D)$ operations. Every path found by QueryPath($\mathcal{D}$) has at most $n/D$ nodes, since the longest path length in $G$ is bounded by $n/D$ (refer to Section \ref{sec:path-length-red}). Therefore, $\Oish \left( n^2/D^2 \right)$ bits are communicated across any sequence of QueryPath($\mathcal{D}$) operations.
\end{proof}

\begin{proof}[Proof of Lemma~\ref{lem:datastructure}]
Lemma~\ref{lem:ds-preprocessing} initializes the data structure
with all vertices unmarked. Since the labels are known to both
players, they can determine without communication whether some
vertex has label at least $D/2$.
If no such vertex exists, Lemma~\ref{lem:no-long-unmarked-path}
justifies returning \textsc{null}. Otherwise,
Lemma~\ref{lem:find-long-path} finds a path containing at least
$D/2$ unmarked vertices. The players mark these vertices and
maintain the labels using Lemma~\ref{lem:ds-marking}.
Thus the query operation satisfies the requirements of
Lemma~\ref{lem:datastructure}.

Preprocessing uses $\Oish(nD)$ bits by
Lemma~\ref{lem:ds-preprocessing}, and all marking updates
together use $\Oish(nD)$ bits by Lemma~\ref{lem:ds-marking}.
By Lemma~\ref{lem:total-path-finding}, finding all returned paths
uses $\Oish(n^2/D^2)$ bits in total.
Therefore, the total communication is $\Oish(nD+n^2/D^2)$.
\end{proof}

Combining Lemmas~\ref{lem:datastructure}
and~\ref{lem:pathcovcorrectness} proves Lemma~\ref{lem:CCPC}.
Lemma~\ref{lem:CCPathBFS} follows directly from
Lemma~\ref{lem:pathBFS-alg}.

\paragraph*{Disclosure of AI assistance.}
The authors used ChatGPT (OpenAI) for manuscript review and editing. This included reviewing mathematical arguments, assisting with localized proof corrections (primarily in the appendix lemmas), and revising proof exposition. It also assisted with citation checks and LaTeX editing. It was not used to develop the main ideas behind the presented results. The authors take full responsibility for the manuscript and its results.

\clearpage 
\bibliography{refs.bib}

@article{BBBDFN25,
  title={Greedy Algorithms for Shortcut Sets and Hopsets},
  author={Bals, Ben and Blikstad, Joakim and Bodwin, Greg and Dadush, Daniel and Forster, Sebastian and Nazari, Yasamin},
  journal={arXiv preprint arXiv:2511.20111},
  year={2025}
}

@inproceedings{Yao1979CommunicationComplexity,
  author    = {Yao, Andrew Chi-Chih},
  title     = {Some Complexity Questions Related to Distributive Computing},
  booktitle = {Proceedings of the Eleventh Annual ACM Symposium on Theory of Computing},
  series    = {STOC '79},
  pages     = {209--213},
  year      = {1979},
  publisher = {Association for Computing Machinery}
}

@book{kushilevitz1997communication,
  author    = {Eyal Kushilevitz and Noam Nisan},
  title     = {Communication Complexity},
  publisher = {Cambridge University Press},
  year      = {1997}
}

@article{kalyanasundaram1992probabilistic,
  author  = {Bala Kalyanasundaram and Georg Schnitger},
  title   = {The Probabilistic Communication Complexity of Set Intersection},
  journal = {SIAM Journal on Discrete Mathematics},
  volume  = {5},
  number  = {4},
  pages   = {545--557},
  year    = {1992}
}

@article{razborov1992distributional,
  author  = {Alexander A. Razborov},
  title   = {On the Distributional Complexity of Disjointness},
  journal = {Theoretical Computer Science},
  volume  = {106},
  number  = {2},
  pages   = {385--390},
  year    = {1992}
}

@article{bar2004information,
  author  = {Ziv Bar-Yossef and T. S. Jayram and Ravi Kumar and D. Sivakumar},
  title   = {An Information Statistics Approach to Data Stream and Communication Complexity},
  journal = {Journal of Computer and System Sciences},
  volume  = {68},
  number  = {4},
  pages   = {702--732},
  year    = {2004}
}

@article{chattopadhyay2010set,
  author  = {Arkadev Chattopadhyay and Toniann Pitassi},
  title   = {The Story of Set Disjointness},
  journal = {SIGACT News},
  volume  = {41},
  number  = {3},
  pages   = {59--85},
  year    = {2010}
}

@inproceedings{weinstein2015simultaneous,
  author    = {Omri Weinstein},
  title     = {The Simultaneous Communication of Disjointness with Applications to Data Streams},
  booktitle = {Proceedings of the 32nd International Symposium on Theoretical Aspects of Computer Science (STACS)},
  pages     = {108--120},
  year      = {2015}
}

@incollection{lu_better_2022,
  title     = {Better {Lower} {Bounds} for {Shortcut} {Sets} and {Additive} {Spanners} via an {Improved} {Alternation} {Product}},
  url       = {https://epubs.siam.org/doi/abs/10.1137/1.9781611977073.131},
  booktitle = {Proceedings of the 2022 {Annual} {ACM}-{SIAM} {Symposium} on {Discrete} {Algorithms} ({SODA})},
  author    = {Lu, Kevin and Williams, Virginia Vassilevska and Wein, Nicole and Xu, Zixuan},
  year      = {2022},
  doi       = {10.1137/1.9781611977073.131},
  note      = {\_eprint: https://epubs.siam.org/doi/pdf/10.1137/1.9781611977073.131},
  pages     = {3311--3331}
}

@inproceedings{kogan_beating_2022,
  title        = {Beating {Matrix} {Multiplication} for $n^{1/3}$-{Directed} {Shortcuts}},
  volume       = {229},
  copyright    = {Creative Commons Attribution 4.0 International license, info:eu-repo/semantics/openAccess},
  issn         = {1868-8969},
  url          = {https://drops.dagstuhl.de/entities/document/10.4230/LIPIcs.ICALP.2022.82},
  doi          = {10.4230/LIPICS.ICALP.2022.82},
  language     = {en},
  urldate      = {2024-11-04},
  booktitle    = {49th International Colloquium on Automata, Languages, and Programming (ICALP 2022)},
  series       = {Leibniz International Proceedings in Informatics (LIPIcs)},
  publisher    = {Schloss Dagstuhl -- Leibniz-Zentrum f{\"u}r Informatik},
  author       = {Kogan, Shimon and Parter, Merav},
  collaborator = {Bojańczyk, Mikołaj and Merelli, Emanuela and Woodruff, David P.},
  year         = {2022},
  source_metadata = {Artwork Size: 20 pages, 1031574 bytes
                  ISBN: 9783959772358
                  Medium: application/pdf
                  Publisher: Schloss Dagstuhl – Leibniz-Zentrum für Informatik},
  pages        = {82:1--82:20}
}

@inproceedings{cao_efficient_2020,
  address   = {New York, NY, USA},
  series    = {{STOC} 2020},
  title     = {Efficient construction of directed hopsets and parallel approximate shortest paths},
  isbn      = {978-1-4503-6979-4},
  url       = {https://doi.org/10.1145/3357713.3384270},
  doi       = {10.1145/3357713.3384270},
  booktitle = {Proceedings of the 52nd {Annual} {ACM} {SIGACT} {Symposium} on {Theory} of {Computing}},
  publisher = {Association for Computing Machinery},
  author    = {Cao, Nairen and Fineman, Jeremy T. and Russell, Katina},
  year      = {2020},
  note      = {event-place: Chicago, IL, USA},
  pages     = {336--349}
}

@article{huang_lower_2021,
  title    = {Lower {Bounds} on {Sparse} {Spanners}, {Emulators}, and {Diameter}-{Reducing} {Shortcuts}},
  volume   = {35},
  issn     = {0895-4801, 1095-7146},
  url      = {https://epubs.siam.org/doi/10.1137/19M1306154},
  doi      = {10.1137/19M1306154},
  language = {en},
  number   = {3},
  urldate  = {2024-12-19},
  journal  = {SIAM Journal on Discrete Mathematics},
  author   = {Huang, Shang-En and Pettie, Seth},
  month    = jan,
  year     = {2021},
  pages    = {2129--2144}
}

@article{berman_finding_nodate,
        title = {Finding {Sparser} {Directed} {Spanners}},
        language = {en},
        author = {Berman, Piotr and Raskhodnikova, Sofya and Ruan, Ge},
}

@InProceedings{FWY20,
  author =	{Manuel Fern{\'a}ndez V and David P. Woodruff and Taisuke Yasuda},
  title =	{{Graph Spanners in the Message-Passing Model}},
  booktitle =	{11th Innovations in Theoretical Computer Science Conference (ITCS 2020)},
  pages =	{77:1--77:18},
  series =	{Leibniz International Proceedings in Informatics (LIPIcs)},
  ISBN =	{978-3-95977-134-4},
  ISSN =	{1868-8969},
  year =	{2020},
  volume =	{151},
  editor =	{Thomas Vidick},
  publisher =	{Schloss Dagstuhl--Leibniz-Zentrum fuer Informatik},
  address =	{Dagstuhl, Germany},
  URL =		{https://drops.dagstuhl.de/opus/volltexte/2020/11762},
  URN =		{urn:nbn:de:0030-drops-117620},
  doi =		{10.4230/LIPIcs.ITCS.2020.77}
}

@inproceedings{HXX25,
  title={New Separations and Reductions for Directed Hopsets and Preservers},
  author={Hoppenworth, Gary and Xu, Yinzhan and Xu, Zixuan},
  booktitle={2025 ACM-SIAM Symposium on Discrete Algorithms (SODA)},
  year={2025},
  pages={4405--4443}
}

@inproceedings{BH23,
  title     = {Folklore {Sampling} is {Optimal} for {Exact} {Hopsets}: {Confirming} the $\sqrt{n}$ {Barrier}},
  doi       = {10.1109/FOCS57990.2023.00046},
  booktitle = {2023 {IEEE} 64th {Annual} {Symposium} on {Foundations} of {Computer} {Science} ({FOCS})},
  author    = {Bodwin, Greg and Hoppenworth, Gary},
  year      = {2023},
  organization = {IEEE},
  pages     = {701--720}
}

@inproceedings{JLS19,
  title={Parallel reachability in almost linear work and square root depth},
  author={Jambulapati, Arun and Liu, Yang P and Sidford, Aaron},
  booktitle={2019 IEEE 60th Annual Symposium on Foundations of Computer Science (FOCS)},
  pages={1664--1686},
  year={2019},
  organization={IEEE}
}

@inproceedings{VXX24,
  title={Simpler and higher lower bounds for shortcut sets},
  author={Williams, Virginia Vassilevska and Xu, Yinzhan and Xu, Zixuan},
  booktitle={Proceedings of the 2024 Annual ACM-SIAM Symposium on Discrete Algorithms (SODA)},
  pages={2643--2656},
  year={2024},
  organization={SIAM}
}

@Article{BGJRW12,
  author    = {Bhattacharyya, Arnab and Grigorescu, Elena and Jung, Kyomin and Raskhodnikova, Sofya and Woodruff, David P},
  title     = {Transitive-closure spanners},
  journal   = {SIAM Journal on Computing},
  year      = {2012},
  volume    = {41},
  number    = {6},
  pages     = {1380--1425},
  publisher = {SIAM},
}

@InCollection{Raskhodnikova10,
  author    = {Raskhodnikova, Sofya},
  title     = {Transitive-closure spanners: A survey},
  booktitle = {Property testing},
  year      = {2010},
  publisher = {Springer},
  pages     = {167--196},
}

@InProceedings{HP18,
  author =	{Shang-En Huang and Seth Pettie},
  title =	{{Lower Bounds on Sparse Spanners, Emulators, and Diameter-reducing shortcuts}},
  booktitle =	{16th Scandinavian Symposium and Workshops on Algorithm  Theory (SWAT 2018)},
  pages =	{26:1--26:12},
  series =	{Leibniz International Proceedings in Informatics (LIPIcs)},
  ISBN =	{978-3-95977-068-2},
  ISSN =	{1868-8969},
  year =	{2018},
  volume =	{101},
  editor =	{David Eppstein},
  publisher =	{Schloss Dagstuhl--Leibniz-Zentrum fuer Informatik},
  address =	{Dagstuhl, Germany},
  URL =		{http://drops.dagstuhl.de/opus/volltexte/2018/8852},
  URN =		{urn:nbn:de:0030-drops-88521},
  doi =		{10.4230/LIPIcs.SWAT.2018.26}
}

@article{UY91,
  title={High-probability parallel transitive-closure algorithms},
  author={Ullman, Jeffrey D and Yannakakis, Mihalis},
  journal={SIAM Journal on Computing},
  volume={20},
  number={1},
  pages={100--125},
  year={1991},
  publisher={SIAM}
}

@inproceedings{KP22a,
  title={New Diameter-Reducing Shortcuts and Directed Hopsets: Breaking the Barrier},
  author={Kogan, Shimon and Parter, Merav},
  booktitle={Proceedings of the 2022 Annual ACM-SIAM Symposium on Discrete Algorithms (SODA)},
  pages={1326--1341},
  year={2022},
  organization={SIAM}
}

@misc{haeupler2026certified,
  author = {Bernhard Haeupler and Antti Roeyskoe and Zhijun Zhang},
  title  = {Better Diameter Bounds for Efficient Shortcuts and a Structural Criterion for Constructiveness},
  year   = {2026},
  eprint = {2602.10747},
  archivePrefix = {arXiv},
  primaryClass = {cs.DS}
}

@article{Fineman19,
  title={Nearly work-efficient parallel algorithm for digraph reachability},
  author={Fineman, Jeremy T},
  journal={SIAM Journal on Computing},
  volume={49},
  number={5},
  pages={STOC18--500},
  year={2019},
  publisher={SIAM}
}

@inproceedings{BHT23,
  title={Bridge girth: A unifying notion in network design},
  author={Bodwin, Greg and Hoppenworth, Gary and Trabelsi, Ohad},
  booktitle={2023 IEEE 64th Annual Symposium on Foundations of Computer Science (FOCS)},
  pages={600--648},
  year={2023},
  organization={IEEE}
}

@InProceedings{Hesse03,
  author       = {Hesse, William},
  title        = {Directed graphs requiring large numbers of shortcuts},
  booktitle    = {Proceedings of the fourteenth annual ACM-SIAM symposium on Discrete algorithms},
  year         = {2003},
  organization = {Society for Industrial and Applied Mathematics},
  pages        = {665--669},
}

@InProceedings{Thorup92,
  author       = {Thorup, Mikkel},
  title        = {On shortcutting digraphs},
  booktitle    = {International Workshop on Graph-Theoretic Concepts in Computer Science},
  year         = {1992},
  organization = {Springer},
  pages        = {205--211},
}

@inproceedings{BRR10,
  title={Finding sparser directed spanners},
  author={Berman, Piotr and Raskhodnikova, Sofya and Ruan, Ge},
  booktitle={IARCS Annual Conference on Foundations of Software Technology and Theoretical Computer Science (FSTTCS 2010)},
  pages={424--435},
  year={2010},
  organization={Schloss Dagstuhl--Leibniz-Zentrum fuer Informatik}
}

@inproceedings{Elkin04,
  title={Unconditional lower bounds on the time-approximation tradeoffs for the distributed minimum spanning tree problem},
  author={Elkin, Michael},
  booktitle={Proceedings of the thirty-sixth annual ACM symposium on Theory of computing},
  pages={331--340},
  year={2004}
}

@article{PR00,
  title={A near-tight lower bound on the time complexity of distributed minimum-weight spanning tree construction},
  author={Peleg, David and Rubinovich, Vitaly},
  journal={SIAM Journal on Computing},
  volume={30},
  number={5},
  pages={1427--1442},
  year={2000},
  publisher={SIAM}
}

@inproceedings{NDP11,
  title={A tight unconditional lower bound on distributed randomwalk computation},
  author={Nanongkai, Danupon and Das Sarma, Atish and Pandurangan, Gopal},
  booktitle={Proceedings of the 30th annual ACM SIGACT-SIGOPS symposium on Principles of distributed computing},
  pages={257--266},
  year={2011}
}

@inproceedings{Nanongkai14,
  title={Distributed approximation algorithms for weighted shortest paths},
  author={Nanongkai, Danupon},
  booktitle={Proceedings of the forty-sixth annual ACM symposium on Theory of computing},
  pages={565--573},
  year={2014}
}

@inproceedings{HW12,
  title={Optimal distributed all pairs shortest paths and applications},
  author={Holzer, Stephan and Wattenhofer, Roger},
  booktitle={Proceedings of the 2012 ACM symposium on Principles of distributed computing},
  pages={355--364},
  year={2012}
}

@inproceedings{FHW12,
  title={Networks cannot compute their diameter in sublinear time},
  author={Frischknecht, Silvio and Holzer, Stephan and Wattenhofer, Roger},
  booktitle={Proceedings of the twenty-third annual ACM-SIAM symposium on Discrete Algorithms},
  pages={1150--1162},
  year={2012},
  organization={SIAM}
}

@inproceedings{DKO14,
  title={On the power of the congested clique model},
  author={Drucker, Andrew and Kuhn, Fabian and Oshman, Rotem},
  booktitle={Proceedings of the 2014 ACM symposium on Principles of distributed computing},
  pages={367--376},
  year={2014}
}

@inproceedings{DHKKNPPW11,
  title={Distributed verification and hardness of distributed approximation},
  author={Das Sarma, Atish and Holzer, Stephan and Kor, Liah and Korman, Amos and Nanongkai, Danupon and Pandurangan, Gopal and Peleg, David and Wattenhofer, Roger},
  booktitle={Proceedings of the forty-third annual ACM symposium on Theory of computing},
  pages={363--372},
  year={2011}
}

@article{CKPY18,
  title={Distributed construction of purely additive spanners},
  author={Censor-Hillel, Keren and Kavitha, Telikepalli and Paz, Ami and Yehudayoff, Amir},
  journal={Distributed Computing},
  volume={31},
  number={3},
  pages={223--240},
  year={2018},
  publisher={Springer}
}

@inproceedings{BCDELP19,
  title={Hardness of distributed optimization},
  author={Bacrach, Nir and Censor-Hillel, Keren and Dory, Michal and Efron, Yuval and Leitersdorf, Dean and Paz, Ami},
  booktitle={Proceedings of the 2019 ACM Symposium on Principles of Distributed Computing},
  pages={238--247},
  year={2019}
}

@article{Roughgarden16,
  title={Communication complexity (for algorithm designers)},
  author={Roughgarden, Tim and others},
  journal={Foundations and Trends{\textregistered} in Theoretical Computer Science},
  volume={11},
  number={3--4},
  pages={217--404},
  year={2016},
  publisher={Now Publishers, Inc.}
}

@inproceedings{LN2022,
  title={Near-Optimal Decremental Hopsets with Applications},
  author={{\L}{\k{a}}cki, Jakub and Nazari, Yasamin},
  booktitle={49th International Colloquium on Automata, Languages, and Programming (ICALP 2022)},
  year={2022},
  organization={Schloss Dagstuhl-Leibniz-Zentrum f{\"u}r Informatik}
}

@inproceedings{bernstein2021deterministic,
  title={Deterministic Decremental {SSSP} and Approximate Min-Cost Flow in Almost-Linear Time},
  author={Bernstein, Aaron and Probst Gutenberg, Maximilian and Saranurak, Thatchaphol},
  booktitle={62 Annual IEEE Symposium on Foundatios of Computer Science (FOCS 2022)},
  year={2021}
}

@inproceedings{chen2022maximum,
  title={Maximum flow and minimum-cost flow in almost-linear time},
  author={Chen, Li and Kyng, Rasmus and Liu, Yang P and Peng, Richard and Gutenberg, Maximilian Probst and Sachdeva, Sushant},
  booktitle={2022 IEEE 63rd Annual Symposium on Foundations of Computer Science (FOCS)},
  pages={612--623},
  year={2022},
  organization={IEEE}
}

@inproceedings{cao2023exact,
  title={Parallel exact shortest paths in almost linear work and square root depth},
  author={Cao, Nairen and Fineman, Jeremy T},
  booktitle={Proceedings of the 2023 Annual ACM-SIAM Symposium on Discrete Algorithms (SODA)},
  pages={4354--4372},
  year={2023},
  organization={SIAM}
}

@inproceedings{cao2020improved,
  title={Improved Work Span Tradeoff for Single Source Reachability and Approximate Shortest Paths},
  author={Cao, Nairen and Fineman, Jeremy T and Russell, Katina},
  booktitle={ACM Symposium on Parallelism in Algorithms and Architectures},
  year={2020}
}

@inproceedings{fineman2018nearly,
  title={Nearly work-efficient parallel algorithm for digraph reachability},
  author={Fineman, Jeremy T},
  booktitle={Proceedings of the 50th Annual ACM SIGACT Symposium on Theory of Computing},
  pages={457--470},
  year={2018}
}

@inproceedings{elkin2019RNC,
  title={Linear-size hopsets with small hopbound, and constant-hopbound hopsets in {RNC}},
  author={Elkin, Michael and Neiman, Ofer},
  booktitle={The 31st ACM Symposium on Parallelism in Algorithms and Architectures},
  pages={333--341},
  year={2019}
}

@article{elkin2019journal,
  title={Hopsets with constant hopbound, and applications to approximate shortest paths},
  author={Elkin, Michael and Neiman, Ofer},
  journal={SIAM Journal on Computing},
  year={2019},
  publisher={SIAM}
}

@inproceedings{elkin2017,
  title={Near-Optimal Distributed Routing with Low Memory},
  author={Elkin, Michael and Neiman, Ofer},
  booktitle={Proceedings of the ACM Symposium on Principles of Distributed Computing},
  year={2018},
  organization={ACM}
}

@article{censor2021,
  title={Fast approximate shortest paths in the congested clique},
  author={Censor-Hillel, Keren and Dory, Michal and Korhonen, Janne H and Leitersdorf, Dean},
  journal={Distributed Computing},
  volume={34},
  pages={463--487},
  year={2021},
  publisher={Springer}
}

@article{cohen2000,
  title={Polylog-time and near-linear work approximation scheme for undirected shortest paths},
  author={Cohen, Edith},
  journal={Journal of the ACM (JACM)},
  year={2000},
  publisher={ACM}
}

@inproceedings{henzinger2014,
  title={Decremental single-source shortest paths on undirected graphs in near-linear total update time},
  author={Henzinger, Monika and Krinninger, Sebastian and Nanongkai, Danupon},
  booktitle={2014 IEEE 55th Annual Symposium on Foundations of Computer Science},
  pages={146--155},
  year={2014},
  organization={IEEE}
}

@inproceedings{gutenberg2020,
  title={Deterministic algorithms for decremental approximate shortest paths: Faster and simpler},
  author={Gutenberg, Maximilian Probst and Wulff-Nilsen, Christian},
  booktitle={Proceedings of the Fourteenth Annual ACM-SIAM Symposium on Discrete Algorithms},
  pages={2522--2541},
  year={2020},
  organization={SIAM}
}

\clearpage 
\appendix
\section{Lower Bound}
In this section, we focus on lower bounding the amount of communication required to compute a $\Theta(D)$-diameter reducing shortcut set for a fixed parameter $D$. This is done by embedding a large instance of \textsc{SetDisj} in a $\infty$-bridge-free path system.

\begin{itemize}
    \item  A \emph{path system} is a pair $\mathcal S=(V,\Pi)$, where $V$ is a set of vertices and $\Pi$ is a multiset of simple vertex sequences called paths; its size is $\|\mathcal S\|=\sum_{\pi\in\Pi}|\pi|$. 
    \item A \emph{$b$-bridge} in $\mathcal S$ consists of distinct vertices $v_1,\ldots,v_b$ and distinct paths $\pi_1,\ldots,\pi_b$ such that $v_i$ precedes $v_{i+1}$ on $\pi_i$ for every $i<b$, while $v_1$ precedes $v_b$ on $\pi_b$; $\pi_b$ is called the \emph{river} and the other paths are \emph{arcs}. 
    \item The \emph{bridge girth} of $\mathcal S$ is the minimum $b$ for which a $b$-bridge exists, or $\infty$ if no bridge exists. Thus, $\mathcal S$ is \emph{$b$-bridge-free} if its bridge girth is greater than $b$. 
    \item Finally, $\beta(n,p,b)$ denotes the maximum possible size $\|\mathcal S\|$ of a path system on $n$ vertices and $p$ paths with bridge girth greater than $b$.
\end{itemize}


We use the following consequence of the known lower bounds for infinite
bridge girth.  Recall that the size of a path system
$S=(V,\Pi)$ is $\lVert S\rVert=\sum_{\pi\in\Pi}|\pi|$.

\begin{lemma}[Path-system input lemma]\label{lem:path-system-input}
Fix an integer $c\geq 1$ and a constant $\epsilon>0$.
There is a constant $\delta=\delta(c,\epsilon)>0$ such that,
for all sufficiently large $n$ and every integer $1\leq D\leq\delta n$,
there is an acyclic, $\infty$-bridge-free path system $S=(V,\Pi)$
on at most $n$ vertices satisfying
\[
  m:=\sum_{\pi\in\Pi}(|\pi|-1)
  =\Omega\!\left(\frac{n^{2c/(c+1)}}{D^{c-1}}\right),
  \qquad pD\leq\epsilon m,
  \qquad p:=|\Pi|.
\]
\end{lemma}

\begin{proof}
We may assume $\epsilon\leq 1$.
Corollaries 41 and 42 of \cite{BHT23}, together with the single $n$-vertex path,
give a constant $\kappa=\kappa(c)>0$ such that, for every integer $P\geq 1$,
there is such a system with at most $P$ nonempty paths and size
\[
  M:=\sum_{\pi\in\Pi}|\pi|
  \geq\kappa\left(n+n^{2/(c+1)}P^{(c-1)/c}\right).
\]
Choose $0<\delta\leq\kappa\epsilon/2$, and set
\[
  F:=\frac{n^{2c/(c+1)}}{D^{c-1}},
  \qquad
  P:=\left\lceil\frac{\delta^cF}{D}\right\rceil.
\]
The resulting system satisfies $M\geq\kappa(n+\delta^{c-1}F)$ and
\[
  pD\leq PD\leq D+\delta^cF
  \leq\delta(n+\delta^{c-1}F)
  \leq\frac{\epsilon}{2}M.
\]
Since $D\geq 1$ and $\epsilon\leq 1$, we have $p\leq M/2$.
Thus $m=M-p\geq M/2=\Omega(F)$ and $pD\leq\epsilon m$.
\end{proof}

\begin{theorem}[Communication lower bound]\label{thm:shortcut-lower-bound}
Fix an integer $c\geq 1$ and an error bound $0\leq\rho<1/2$.
There is a constant $\delta=\delta(c,\rho)>0$ such that, for all
sufficiently large $n$ and every integer $1\leq D\leq\delta n$,
every two-party protocol that outputs a correct $D$-diameter-reducing
shortcut set with error at most $\rho$ has worst-case communication cost
\[
  \Omega\!\left(\frac{n^{2c/(c+1)}}{D^{c-1}}\right).
\]
Here Alice and Bob receive edge sets $E_A,E_B$ defining
$G=(V,E_A\cup E_B)$, the sets may overlap, and both players agree on a shortcut set.
The bound holds even for DAGs, with no restriction on the shortcut-set size.
\end{theorem}

\begin{proof}
Let $a=a(\rho)>0$ satisfy
$R_\rho(\operatorname{DISJ}_m)\geq am$.
Apply Lemma~\ref{lem:path-system-input} with a sufficiently small constant
$\epsilon>0$, to be chosen below.  Let $G_S=(V,E_S)$ be the union of the
path edges, padding with isolated vertices to obtain $n$ vertices.
Infinite bridge girth implies that the paths are edge-disjoint and that
every forward subpath $\pi[u,v]$ is the unique directed $u\leadsto v$ path
in $G_S$; see \cite[proof of Lemma~40]{BHT23}.
In particular, $|E_S|=m$.

The path system $S$ and the ordering of $E_S$ are fixed and known to both players, independent of their vectors in the $\operatorname{DISJ}_m$ game or their edge sets in the shortcut set game.

Index an instance $x,y\in\{0,1\}^{E_S}$ of
$\operatorname{DISJ}_m$ by these edges, and define
\[
  E_A:=\{e\in E_S:x_e=0\},
  \qquad
  E_B:=\{e\in E_S:y_e=0\}.
\]
Thus an edge $e\in E_S$ is missing from $G=(V,E_A\cup E_B)$ exactly
when $x_e=y_e=1$.

Run any shortcut set protocol that uses $q$ bits.
Each player uses their own output as $H$ below; when the protocol succeeds, these outputs coincide.

For each $\pi\in\Pi$, let $U_\pi$ be its edges not spanned by any shortcut edge:
\[
  U_\pi:=E(\pi)\setminus
  \bigcup_{\substack{(u,v)\in H\\u,v\in\pi,\ u<_\pi v}}
  E(\pi[u,v]).
\]
Whenever the shortcut protocol succeeds, the following hold.
\begin{enumerate}
\item Every edge in $E(\pi)\setminus U_\pi$ is present in $G$. This is because every edge $(u,v)$ in the shortcut set $H$ is from the transitive closure of $G$, so it certifies a $u \rightsquigarrow v$ path in $G$. Additionally, if the shortcut edge endpoints $u,v$ are found along a path $\pi \in S$, then the subpath $\pi[u,v]$ is the unique $u$ to $v$ path in $G$, so all edges along $\pi[u,v]$ must be present in $G$.
\item If every edge of $\pi$ is present in $G$ then $|U_\pi|\leq D$. This is because there is a unique path for each pair of endpoints and $H$ reduces the diameter of the graph to below $D$. Thus, the path between the endpoints of $\pi$ in $G \cup H$ would need to use all the edges in $U_{\pi}$ and have at most $D$ edges.
\end{enumerate}

For $P\in\{A,B\}$, let $U_\pi^P$ be the set computed from player $P$'s output, and let $U^P:=\bigcup_{\pi\in\Pi}U_\pi^P$.
Alice sends the bits $(x_e)_{e\in U^A}$ in the public edge order, truncating or padding with zeros to exactly $pD$ bits.
If Bob finds some $\pi$ with $|U_\pi^B|>D$, he answers ``not disjoint.''
Otherwise, $|U^B|\le pD$, and Bob compares the first $|U^B|$ received bits with $(y_e)_{e\in U^B}$ in the public edge order. 
He answers ``disjoint'' exactly when for every bit index $i$, at least one of the entries at $i$ is $0$.

Choose $\epsilon$ small enough that $\epsilon m$ is at most $am/2$ for
sufficiently large $n$.
The resulting disjointness protocol has error at most $\rho$, because
its postprocessing is correct whenever $H$ is correct.
Consequently, $q+am/2\geq am$, so $q\geq am/2$.
The claimed bound follows from Lemma~\ref{lem:path-system-input}.
\end{proof}

\section{Simplifying Reductions in the Upper Bound} \label{sec:reds}

\subsection{A low-communication reduction to DAGs}

\begin{lemma}[SCC contraction]\label{lem:scc-contraction}
Given a directed graph $G=G_A\cup G_B$ on $n$ vertices, Alice and Bob can
identify all strongly connected components using $\widetilde O(n)$ bits of
communication.  Once the SCC partition is known, they can agree without further
communication on $O(n)$ shortcut edges that give hop diameter at most $2$
inside every strongly connected component.
\end{lemma}

\begin{proof}
The players maintain a current graph obtained by contracting already
identified strongly connected sets, a stack of current vertices, and a set of
finished vertices.  The stack is ordered from bottom to top.  Whenever a set
is contracted, each player maps the endpoints of her or his local edges to the
new supervertex and discards self-loops.

If the stack is empty, the players push any unfinished vertex.  Otherwise,
each player searches her or his edges for an \emph{eligible} edge $(x,y)$:
$x$ lies on the stack, $y$ is unfinished, and either $y$ is not on the stack
or $y$ occurs strictly below $x$ on the stack.  A player who has an eligible
edge sends one whose tail is highest on the stack; a player with no such edge
sends \textsc{null}.  If at least one edge is sent, the players select the sent
edge whose tail is highest, say $(x,y)$, and delete from the stack every
vertex strictly above $x$, declaring each deleted current vertex finished.

If $y$ is not on the stack, they push $y$.  If $y$ is already on the stack,
the stack segment from $y$ to $x$ is a directed path of already communicated
edges.  Together with $(x,y)$ it is a directed cycle, so the players contract
the entire segment to one supervertex.  If both players send \textsc{null},
they declare every vertex currently on the stack finished and clear the
stack.

We prove correctness using two invariants.

\begin{enumerate}
    \item[(I1)] The stack, in its stack order, is a directed path of known edges.
    \item[(I2)] No edge of the current graph leaves the set of finished vertices.
\end{enumerate}

(I1) holds initially.  Pushing a vertex appends the communicated edge, and
contracting a stack interval replaces a directed subpath by one vertex.

For (I2), consider vertices deleted from the top of the stack after $(x,y)$
is selected.  None of them has an eligible outgoing edge: otherwise the owner
of such an edge would have sent an edge with tail above $x$.  Hence no newly
finished vertex has an edge to an unfinished vertex off the deleted suffix or
to a vertex remaining lower on the stack.  Edges to previously finished
vertices or within the deleted suffix do not violate (I2).  The same argument
applies when the whole stack is cleared after two \textsc{null} messages.
Thus (I2) is preserved.

Every contraction is safe because it contracts a known directed cycle and
hence a strongly connected set.  Conversely, no two distinct current
vertices belonging to the same strongly connected component can become
finished.  To see this, consider the first round in which this could happen.
They cannot be separated between the old finished set and its complement by
(I2).  They also cannot be two distinct vertices in a newly finished stack
suffix: a directed cycle inside that suffix would contain an edge from a
higher stack vertex to a lower one, which would be eligible, while a cycle
leaving the suffix would first use an eligible edge to an unfinished vertex
off the suffix.  A path through the old finished set cannot return by (I2).
The same reasoning applies when the entire stack is cleared.  Therefore all
vertices in each maximal strongly connected component are contracted
together, and distinct components are never merged.

Each non-\textsc{null} message specifies two vertex names and costs
$O(\log n)$ bits.  A vertex is first pushed at most once, a finished vertex
never returns to the stack, and every contraction reduces the number of
current vertices.  Thus there are $O(n)$ pushes, $O(n)$ deletions, and at most
$n-1$ contractions.  There are consequently $O(n)$ rounds and
$\widetilde O(n)$ communicated bits.

Finally, after the maximal SCCs are known, choose a canonical representative
$r_K$ for each SCC $K$ and add
\[
    (r_K,v),(v,r_K)\qquad (v\in K\setminus\{r_K\}).
\]
These are valid transitive-closure edges, use fewer than $2n$ edges in total,
and give diameter at most $2$ inside each SCC.  No communication is needed
because the SCC partition and the choice of representatives are common
knowledge.
\end{proof}

\subsubsection{Lifting shortcut edges through the condensation}
Let $\overline G$ be the condensation DAG.  A shortcut edge $(K,L)$ in
$\overline G^*$ is lifted canonically to $(r_K,r_L)\in G^*$.  If a shortcut
set gives diameter $D_0$ in $\overline G$, the lifted edges together with the
SCC stars give diameter at most $3D_0+2$ in $G$.  Indeed, a lifted shortcut
edge costs one hop, while an ordinary condensation edge $K\to L$ can be
realized by going from $r_K$ to the tail of an original edge from $K$ to
$L$, traversing that edge, and then going to $r_L$, for at most three hops.
There is one additional star edge at each endpoint.  Thus, to obtain an exact
target $D \geq 100$, run the DAG protocol with target
$D_0=\lfloor(D-2)/3\rfloor$.  This constant-factor change affects none of the
asymptotic size or communication bounds.
For $1\le D<100$, the players exchange their adjacency matrices
and output $E(G^*)$, achieving diameter at most $1$ using
$O(n^2)$ bits and $O(n^2)$ edges. Since $D<100$, these satisfy
the claimed communication and shortcut-size bounds.

\subsection{Reduction to a DAG with bounded longest path} \label{sec:path-length-red}

The following proof uses the capped-label protocol proved in
Section~\ref{sec:labels-from-scratch}.

\begin{lemma}[Bounded-longest-path reduction]\label{lem:bounded-path-reduction}
    Let $G=G_A\cup G_B$ be an $n$-vertex DAG and let $1\le D\le n$.
    Using $\widetilde O(nD)$ bits of communication, Alice and Bob can agree on a collection $C$ of at most $D$ vertex-disjoint paths in $G^*$ and a shortcut set $X\subseteq G^*$ such that
    \[
        G'=(G\cup X)\setminus V(C)
    \]
    has the same reachability relation as $G$ on the remaining vertices, and every directed path in $G'$ has fewer than $\lceil n/D\rceil$ vertices.

For each edge in $X$, the players also know a canonical witnessing
path. Recursively expanding previously added edges of $X$ turns
each witness into a path in $G$ that intersects $V(C)$.
\end{lemma}

\begin{proof}
Set $L=\lceil n/D\rceil$ and $G_0=G$. At iteration $j$, stop if
$G_{j-1}$ has fewer than $L$ vertices. Otherwise, compute the capped
labels from Lemma~\ref{lem:capped-labels} with threshold $L$.
If all labels are below $L$, stop. Otherwise, follow maximizing
successors to recover a directed path $\pi_j$ with exactly $L$
vertices, and compute
\[
    T_j=\PathBFS(\pi_j,G_{j-1}).
\]
Let $X_j$ contain every edge $(a,b)$ with
$a,b\in V(G_{j-1})\setminus V(\pi_j)$ for which $T_j$ contains an
$a\rightsquigarrow b$ path intersecting $\pi_j$. Choose such a path
as a witness using a fixed public rule, retaining the first witness
assigned to an edge if it appears in multiple iterations. Set
\[
    G_j=(G_{j-1}\cup X_j)\setminus V(\pi_j).
\]
Since $T_j$ is known to both players, computing $X_j$ and choosing
its witnesses requires no further communication.

After $q$ iterations, output
\[
    C=(\pi_1,\ldots,\pi_q),
    \qquad
    X=\bigcup_{j=1}^q X_j.
\]
By construction, $G_q=(G\cup X)\setminus V(C)=G'$.

We first show that each iteration preserves reachability among
the remaining vertices. Since $\pi_j$ is a path in $G_{j-1}$,
we have $T_j\subseteq G_{j-1}$. Thus every edge of $X_j$ lies in
$G_{j-1}^*$, so adding $X_j$ preserves reachability and acyclicity.
Conversely, consider an $a\rightsquigarrow b$ path $P$ in $G_{j-1}$
whose endpoints lie outside $\pi_j$. If $P$ avoids $\pi_j$, it
survives in $G_j$. Otherwise, choose any
$x\in V(P)\cap V(\pi_j)$. The guarantee of $\PathBFS$ gives paths
$a\rightsquigarrow x$ and $x\rightsquigarrow b$ in $T_j$.
Their concatenation is a path because $T_j$ is acyclic. Hence
$(a,b)\in X_j$, and $a$ still reaches $b$ in $G_j$.
Inductively, $G'$ has the same reachability relation as $G$ on
the remaining vertices, every edge of $X$ lies in $G^*$, and
every $\pi_j$ is a path in $G^*$.

Now consider an edge first added at iteration $j$. Its witness
intersects $\pi_j$ and uses only original edges and shortcut edges
added in earlier iterations. Recursively expand the latter using
their fixed witnesses. This process terminates because iteration
indices strictly decrease. The resulting walk uses only edges of
$G$ and still contains a vertex of $\pi_j$. Since $G$ is acyclic,
it is a path. Thus every edge of $X$ has the required witness
intersecting $V(C)$.

The paths in $C$ are vertex-disjoint and each has $L$ vertices.
Consequently, $qL\le n$ and $q\le D$. At termination, either fewer
than $L$ vertices remain or all capped labels are below $L$.
In either case, every directed path in $G'$ has fewer than
$L=\lceil n/D\rceil$ vertices.

Finally, there are at most $D+1$ capped-label computations and
at most $D$ calls to $\PathBFS$, each using $\widetilde O(n)$ bits.
Recovering all selected paths uses
$\widetilde O(\sum_j |V(\pi_j)|)=\widetilde O(n)$ additional bits.
The total communication is therefore $\widetilde O(nD)$.
\end{proof}

\begin{lemma}[Lifting from the reduced DAG]\label{lem:lift-bounded-path}
Assume $D\geq 8$, and let $G,X,C,G'$ be as in
Lemma~\ref{lem:bounded-path-reduction}.  Given a $D$-diameter-reducing
shortcut set $H'$ for $G'$, Alice and Bob can use
$\widetilde O(nD)$ additional bits to compute a set
$H\subseteq G^*$ of $\widetilde O(n)$ edges such that
$H\cup H'$ is a $D$-diameter-reducing shortcut set for $G$, with high
probability.
\end{lemma}

\begin{proof}
First note that the reachability guarantee in Lemma~\ref{lem:bounded-path-reduction} implies
$H'\subseteq G^*$.

For every $\pi\in C$, add to $H$ the standard
$O(|\pi|\log |\pi|)$ transitive-closure edges that reduce the hop diameter of
$\pi$ to at most $2$.  Since the paths in $C$ are vertex-disjoint, these
edges total $O(n\log n)$.

Independently sample a public set
\[
    R\subseteq V,
    \qquad
    |R|=\min\left\{n,\left\lceil c_0\frac{n\log n}{D}\right\rceil\right\},
\]
where $c_0$ is a sufficiently large constant.  For each $\pi\in C$, compute
$T_\pi=\PathBFS(\pi,G)$ in the original graph.  For every $r\in R$:
\begin{itemize}
\item if $r$ reaches a vertex of $\pi$, let $a$ be the first such vertex on
      $\pi$ and add $(r,a)$ to $H$; and
\item if a vertex of $\pi$ reaches $r$, let $b$ be the last such vertex on
      $\pi$ and add $(b,r)$ to $H$.
\end{itemize}
The correctness of $\PathBFS$ lets the players determine these vertices from
$T_\pi$.  Every added edge lies in $G^*$.  There are at most
$2|R||C|=\widetilde O(n)$ connector edges, and the $|C|\leq D$ calls to
$\PathBFS$ use $\widetilde O(nD)$ bits.

It remains to prove the diameter guarantee.  Fix the graph $G$, the sets
$X,C$, and the shortcut set $H'$ before sampling $R$.  For every reachable
ordered pair $(u,v)$, fix a path $P_{u,v}$ in $G\cup H'$ as follows.

If $u,v\notin V(C)$, fix a path $Q_{u,v}$ of at most $D$ edges in
$G'\cup H'$.  Replace each occurrence of an edge $e\in X$ on $Q_{u,v}$ by
the canonical original-graph witness fixed for $e$ in
Lemma~\ref{lem:bounded-path-reduction}; recursively expand earlier $X$-edges
inside that witness.  Leave edges of $H'$ unexpanded.  The resulting path is
$P_{u,v}$.  It is a path in $G\cup H'$ because every expanded edge is
replaced by original edges of $G$: the concatenation cannot repeat a vertex,
since all its edges move forward in a topological ordering of the original
DAG.  Moreover, either $Q_{u,v}$ uses no edge
of $X$, in which case $P_{u,v}=Q_{u,v}$ has at most $D$ edges, or
$P_{u,v}$ contains a vertex of $C$.

If $u\in V(C)$ or $v\in V(C)$, instead fix any canonical $u\rightsquigarrow v$ path in
$G$ and call it $P_{u,v}$.  This path contains a vertex of $C$.  All these
choices are independent of the subsequent sample $R$.

For every fixed $P_{u,v}$ with more than $D$ edges, let
$\operatorname{pref}(P_{u,v})$ and $\operatorname{suff}(P_{u,v})$ be its
first and last $\lfloor D/4\rfloor$ vertices.  A standard hitting-set
calculation and a union bound over at most $n^2$ ordered pairs imply that,
with high probability, $R$ intersects both sets for every such fixed path.
Condition on this event.

If $P_{u,v}$ has at most $D$ edges, it already proves the desired bound.
Otherwise $P_{u,v}$ contains a vertex $x$ on some path $\pi\in C$.  Choose
$r_1\in R\cap\operatorname{pref}(P_{u,v})$ and
$r_2\in R\cap\operatorname{suff}(P_{u,v})$.

Suppose first that $x$ lies outside the prefix and suffix.  Let $a$ be the
first vertex of $\pi$ reachable from $r_1$, and let $b$ be the last vertex of
$\pi$ that reaches $r_2$.  Since $r_1\rightsquigarrow x\rightsquigarrow r_2$ and $x\in\pi$, we
have $a\leq_\pi x\leq_\pi b$.  The construction adds $(r_1,a)$ and
$(b,r_2)$, and the path-shortcut edges give $\dist_{G\cup H}(a,b)\leq2$.
Following the prefix of $P_{u,v}$ to $r_1$, these two connectors, the
shortcut inside $\pi$, and the suffix from $r_2$ gives
\[
   \dist_{G\cup H\cup H'}(u,v)
   \leq \frac D4+1+2+1+\frac D4
   \leq D.
\]
Here a segment of $P_{u,v}$ may use edges of $H'$, which are present in the
final graph; and its reachability statements are valid in $G$ because every
edge of $H'$ lies in $G^*$.

If $x$ lies in the prefix, let $b$ be the last vertex of $\pi$ that reaches
$r_2$.  Then $x\leq_\pi b$, so the prefix from $u$ to $x$, at most two hops
inside $\pi$, the connector $(b,r_2)$, and the suffix from $r_2$ to $v$ use
at most $D/2+3\leq D$ edges.  The case in which $x$ lies in the suffix is
symmetric, using the connector from $r_1$ to the first reachable vertex of
$\pi$.  Thus every reachable pair has distance at most $D$.
\end{proof}

\subsection{Computing capped longest-path labels from scratch}
\label{sec:labels-from-scratch}

\begin{lemma}[Capped labels]\label{lem:capped-labels}
Let $G=G_A\cup G_B$ be an $n$-vertex DAG and let $K\in[1,n]$.  There is a
two-party protocol using $\widetilde O(n)$ bits that computes, for every
$v\in V$, the label
\[
  \ell(v)=\min\left\{K,
    \max_{\pi \text{ a directed path starting at }v}|V(\pi)|
  \right\}.
\]
\end{lemma}

\begin{proof}
A label is either finalized or unfinalized.  Initially all labels are
unfinalized.  For $P\in\{A,B\}$, player $P$ declares a vertex $v$
\emph{$P$-ready} once every out-neighbor $u$ with $(v,u)\in E_P$ has a
finalized label.  At that moment player $P$ sends
\[
   \left(v,\lambda_P(v)\right),
   \qquad
   \lambda_P(v):=\max\bigl(\{\ell(u):(v,u)\in E_P\}\cup\{0\}\bigr).
\]
Each player sends this message for $v$ only once.  Once both messages for $v$
have been received, the players finalize
\[
   \ell(v)=\min\{K,1+\max\{\lambda_A(v),\lambda_B(v)\}\}.
\]

The protocol terminates.  Indeed, the subgraph induced by the currently
unfinalized vertices is a DAG and therefore has a sink $v$.  All outgoing
neighbors of $v$ are already finalized, so $v$ is both $A$-ready and
$B$-ready and is finalized next.

Correctness follows by reverse topological induction.  Once all out-neighbor
labels are correct, the maximum number of vertices on a path starting at $v$
is one plus the maximum corresponding value at an out-neighbor, with the
maximum of the empty set interpreted as zero.  Taking the maximum of the two
players' local maxima gives the maximum over all outgoing edges, and capping
at $K$ gives exactly the displayed label.

Each player sends one vertex name and one integer in $[0,K]$ for each vertex.
Thus $O(n\log n)$ bits are communicated.
\end{proof}

\subsubsection{Use in Lemma~\ref{lem:bounded-path-reduction}}
With $K=\lceil n/D\rceil$, some vertex has label $K$ if and only if the
current DAG contains a path with at least $K$ vertices.  Starting from such a
vertex, the players can repeatedly communicate a maximizing successor and
stop after $K$ vertices.  The recurrence guarantees that a successor exists
until then.  This recovers the required path using
$\widetilde O(K)$ bits.  Thus exact, uncapped longest-path lengths are neither
needed nor computed.

\section*{Composition of the upper-bound reductions}

For completeness, the reductions compose as follows.  First apply
Lemma~\ref{lem:scc-contraction}, replace each SCC by its canonical
representative, and obtain the condensation DAG.  Apply
Lemma~\ref{lem:bounded-path-reduction} to this DAG, run the main shortcut-set
protocol on the resulting bounded-path DAG, and lift it using
Lemma~\ref{lem:lift-bounded-path}.  Finally lift shortcut edges between SCCs
to the canonical representatives and add the SCC stars.  Running the DAG
stages with target $D_0=\lfloor(D-2)/3\rfloor$ absorbs both the at-most-three
hops needed to realize an ordinary condensation edge and the two endpoint
SCC hops.
The auxiliary communication is $\widetilde O(nD)$ and the auxiliary shortcut
sets have $\widetilde O(n)$ edges.  In the regime $D\leq n^{1/3}$ these are
dominated respectively by $\widetilde O(n^2/D^2)$ and
$\widetilde O(n^2/D^3)$.

\end{document}